\documentclass{article}

\usepackage{arxiv}

\usepackage[utf8]{inputenc} 
\usepackage[T1]{fontenc}    
\usepackage{hyperref}       
\usepackage{url}            
\usepackage{booktabs}       
\usepackage{amsfonts}       
\usepackage{nicefrac}       
\usepackage{microtype}      
\usepackage{lipsum}
\usepackage{graphicx}
\usepackage{graphicx}
\usepackage{float}
\usepackage{asymptote}
\usepackage{csquotes}
\usepackage{xcolor}
\usepackage{csquotes}
\usepackage{pgf-pie}
\usepackage{makecell}
\usepackage{subcaption}
\usepackage{cite}
\usepackage{amsmath,amssymb,amsfonts}
\usepackage{algorithmic}
\usepackage{graphicx}
\usepackage{textcomp}
\usepackage{xcolor}
\usepackage{float}
\usepackage{placeins}
\usepackage[utf8]{inputenc} 
\usepackage[T1]{fontenc}    
\usepackage{hyperref}       
\usepackage{url}            
\usepackage{booktabs}       
\usepackage{amsfonts}       
\usepackage{nicefrac}       
\usepackage{microtype}      
\usepackage{lipsum}
\usepackage{graphicx}
\usepackage{graphicx}
\usepackage{float}
\usepackage{multirow}
\usepackage{booktabs}

\definecolor{chipgreen}{HTML}{D4EDDA}
\definecolor{chipgreentxt}{HTML}{155724}
\definecolor{chipred}{HTML}{F8D7DA}
\definecolor{chipredtxt}{HTML}{721C24}
\definecolor{chipamber}{HTML}{FFF3CD}
\definecolor{chipambertxt}{HTML}{856404}
 
\newcommand{\chipyes}[1]{\fcolorbox{chipgreen}{chipgreen}{\parbox{\dimexpr\linewidth-2\fboxsep-2\fboxrule}{\scriptsize\textcolor{chipgreentxt}{\strut #1}}}}
\newcommand{\chipno}[1]{\fcolorbox{chipred}{chipred}{\parbox{\dimexpr\linewidth-2\fboxsep-2\fboxrule}{\scriptsize\textcolor{chipredtxt}{\strut #1}}}}
\newcommand{\chippartial}[1]{\fcolorbox{chipamber}{chipamber}{\parbox{\dimexpr\linewidth-2\fboxsep-2\fboxrule}{\scriptsize\textcolor{chipambertxt}{\strut #1}}}}

\usepackage{orcidlink}

\title{Cloud to Edge: Benchmarking LLM Inference on Hardware-Accelerated Single-Board Computers}

\author{
 Harri Renney \orcidlink{0000-0003-1481-5852}\\
  Kaze Digital \& Data\\
  Kaze Consulting\\
  Cheltenham, GL52 8PX, UK \\
  \texttt{harri@kaze-consulting.com} \\
   \And
 Fouad Trad \orcidlink{0000-0003-2241-8195}\\
  Electrical and Computer Engineering\\
  Lebanese American University\\
  Byblos, Lebanon \\
  \texttt{fouad.trad@lau.edu.lb} \\
  \And
Michael Mattarock \\
  Carnegie Mellon University\\
  Pittsburgh\\
  Pennsylvania\\
  USA\\
  \texttt{mattarock@cmu.edu} \\
  \And
  Jayden Evetts\\
  Kaze Digital \& Data\\
  Kaze Consulting\\
  Cheltenham, GL52 8PX, UK \\
  \texttt{jayden@kaze-consulting.com} \\
   \And
  Zena Wood \orcidlink{0000-0001-8843-9832}\\
  University of Exeter\\
  Exeter, UK\\
  \texttt{Z.M.Wood2@exeter.ac.uk} \\
}

\begin{document}
\maketitle
\begin{abstract}
Large language models (LLMs) are becoming increasingly capable at small parameter scales. At the same time, conventional cloud-centric deployment introduces challenges around data privacy, latency, and cost that are acute in operational technology and defence environments. Advances in model distillation, quantisation, and affordable edge accelerators now make local LLM inference on single-board computers (SBC) feasible, but the high dimensionality of the configuration space makes identifying optimal deployments difficult without structured evaluation. Existing LLM-specific edge benchmarking efforts rely on CPU-only inference, poor coverage of genuine single-board computers, and generic evaluation tasks that lack multi-dimensional assessment of hardware effectiveness. This paper proposes a multi-dimensional benchmarking methodology that jointly evaluates inference performance and hardware efficiency across four IoT-suitable edge platform configurations testing single-board computers with the latest available hardware accelerators. Our results reveal the benefits of using hardware accelerators such as NPUs and GPUs, along with multi-dimensional evaluations quantifying the trade-offs between power efficiency, physical device size and token throughput; offering practical guidance for deploying generative AI in privacy-sensitive and connectivity-limited environments such as unmanned vehicles and portable, ruggedised operations.
\end{abstract}


\section{Introduction} \label{sect:s1}

Since the introduction of the transformer architecture \cite{vaswani2017attention}, large language models (LLMs) have demonstrated remarkable emergent capabilities across reasoning, creative writing, and professional examinations \cite{zhanginterpreting, gomez2023confederacy, brin2023large, katz2024gpt}. However, LLM deployment remains predominantly cloud-centric, with large models accessed via API endpoints as LLM-as-a-Service\cite{liagkou5406285taming}, introducing challenges around data privacy, latency, and dependence on stable connectivity \cite{shanmugarasa2025sok,zheng2025review}.

These limitations are acute in critical industries, defined here as sectors whose networks, systems, and assets are fundamental to public health, economic stability, and national security (e.g., those within energy, health, financial, manufacturing, transportation, or communications sectors). In such operational technology (OT) and defence environments, sensitive data often cannot leave a controlled network perimeter, making localised inference a security imperative. As Internet of Everything deployments across these sectors increasingly demand intelligent local processing, the ability to run LLMs on cost-effective edge hardware becomes essential, from distributed cyber-physical systems to remote satellite ground stations. Three convergent developments have recently changed the picture: (i) distilled small language models in the 1.5B–7B parameter range that retain strong generative capability \cite{grattafiori2024llama, qwen2024qwen2}; (ii) post-training quantisation to INT4/INT8 precision, substantially reducing memory requirements with only modest accuracy loss \cite{dettmers20218, shi2025systematic}; and (iii) a new generation of affordable edge accelerators, such as the Hailo-10H (40 TOPS), NVIDIA Jetson Orin Nano Super (67 TOPS), and integrated NPUs like the M5Stack AX630C, bringing meaningful AI compute to platforms under \$350.
The resulting configuration space spans device type, accelerator, model family, parameter count, quantisation level, and inference runtime, making it difficult to identify optimal configurations without structured evaluation. Existing edge benchmarking tools, such as DeepEdgeBench \cite{baller2021deepedgebenchbenchmarkingdeepneural}, primarily target general problem-solving performance. While more recent frameworks, including LEAF \cite{abdulkadhim2026introducing} and BeDGED \cite{nezami2025descriptor}, expand the evaluation to additional edge-relevant dimensions, they remain limited by reliance on CPU-only or legacy GPU-based LLM inference on IoT-class hardware.

This paper makes three contributions:
\begin{itemize}
    \item A gap analysis of existing edge LLM benchmarking approaches, identifying the absence of multi-dimensional evaluation frameworks that jointly consider inference performance, hardware efficiency, and physical deployment constraints for IoT-suitable single-board computers.
    \item Initial steps toward a multi-dimensional evaluation framework for edge LLM deployment, introducing two composite metrics, throughput density (Tps/m\textsuperscript{3}) and energy per million tokens (MJ/Mtok), that extend conventional single-axis benchmarking to account for physical and energy deployment constraints. These metrics are validated through benchmarking four edge platforms and multiple model families at the 1.5B--3B parameter scale.
    \item An analysis of the trade-offs between token throughput, energy efficiency, and physical device size, with direct applicability to Internet of Everything deployments in critical industries, including distributed cyber networks, unmanned aerial vehicles, and satellite ground stations, along with cybersecurity considerations. 
\end{itemize}

\section{Related Work} \label{sect:s2}

\begin{table*}[!htbp]
\centering
\scriptsize
\caption{Positioning of this work relative to closest prior studies}
\label{tab:related-work}
\renewcommand{\arraystretch}{1.4}
\resizebox{\textwidth}{!}{
\begin{tabular}{l p{2.4cm} p{2.1cm} p{2.1cm} p{2.1cm} p{2.1cm} p{3.0cm}}
\hline
 & \textbf{LEAF \cite{abdulkadhim2026introducing}} 
 & \textbf{BeDGED \cite{nezami2025descriptor}} 
 & \textbf{LLMs-at-Edge \cite{huang2025llms}} 
 & \textbf{Abstreiter et al. \cite{abstreiter2025sometimes}} 
 & \textbf{Tummalapalli et al. \cite{tummalapalli2026llm}} 
 & \textbf{This work} \\
\hline

Edge hardware 
& \chippartial{RPi 4/5, Jetson Nano} 
& \chippartial{RPi 5 cluster} 
& \chipno{Jetson AGX Orin} 
& \chippartial{RPi 5, Jetson Orin Nano} 
& \chippartial{Mobile, RPi 5 + NPU, Laptop GPU} 
& \chipyes{M5Stack, RPi 5, RPi 5 + AI HAT+, Jetson Orin} \\

Edge accelerators 
& \chippartial{Maxwell GPU} 
& \chipno{None (CPU only)} 
& \chippartial{Ampere GPU} 
& \chippartial{GPU only} 
& \chipyes{GPU + NPU} 
& \chipyes{GPU + NPU} \\

Inference runtime 
& \chippartial{Ollama} 
& \chippartial{Ollama} 
& \chippartial{Ollama} 
& \chippartial{llama.cpp} 
& \chippartial{MLC, MLX, vLLM} 
& \chippartial{Ollama variants + StackFlow} \\

Metrics 
& \chipyes{Latency, accuracy, Energy-per-Token} 
& \chippartial{Throughput, latency, memory} 
& \chippartial{Throughput, energy, accuracy} 
& \chipyes{Performance, energy, micro-arch} 
& \chippartial{Performance, power, thermal} 
& \chipyes{Throughput, TTFT, energy (MJ/Mtok), throughput density (Tps/m\textsuperscript{3})} \\

\hline
\end{tabular}
}
\end{table*}

Edge AI is a well-established field, with comprehensive taxonomies covering infrastructure, resource management, and scheduling for conventional workloads \cite{gill2025edge}, and general edge AI benchmarking frameworks such as DeepEdgeBench \cite{baller2021deepedgebenchbenchmarkingdeepneural} standardising evaluation on constrained hardware. However, these tools and taxonomies do not address LLM inference, and DeepEdgeBench specifically does not account for the generation dynamics of LLMs \cite{abdulkadhim2026introducing}.

\subsection{Quantisation and Small Models for Edge Deployment}

The feasibility of edge LLM inference rests on two parallel advances. Distilled architectures at 1.5B to 7B parameters, including Llama 3.2, Qwen 2.5, Phi-3.5/4-mini, and Gemma 2/3, now retain strong generative capability through knowledge distillation and synthetic data training \cite{grattafiori2024llama, qwen2024qwen2}. On the compression side, post-training quantisation to INT4/INT8 reduces memory and bandwidth requirements substantially with modest accuracy loss \cite{dettmers20218, shi2025systematic}. The GGUF format and Q4\_K\_M quantisation scheme have become a practical standard for portable deployment across ARM CPUs and NVIDIA GPUs, with Ollama applying Q4\_K\_M by default.
Reports demonstrate 4×–16× model size reduction when compressing FP16 to INT4 (i.e., 75–93.75\% size reduction) \cite{picovoice2026Sub-4-Bit}, while hardware evaluations further show 57–61\% reductions in area and power consumption, demonstrating practical efficiency gains \cite{lee2025qrazor} that greatly benefit edge computing. Sparse quantisation and mixed-precision inference may provide further efficiency gains, although their benefits depend on runtime and accelerator-level support; the present study therefore uses broadly supported uniform 4-bit quantisation as its reproducible baseline.

\subsection{Benchmarking LLMs at the Edge}
 
Two recent frameworks begin to address this gap. LEAF \cite{abdulkadhim2026introducing} assesses LLMs and introduces sustainability metrics (Circular Economy Score, Energy-per-Token) alongside latency and accuracy, testing 4-bit quantised models via Ollama on only two Single-Board Computers (SBCs), a Raspberry Pi 4/5 and a Jetson Nano, alongside desktops with an NVIDIA T400 and other legacy GPUs. BeDGED \cite{nezami2025descriptor} deploys a Raspberry Pi cluster in a small base station (SBS) architecture with lightweight Kubernetes, capturing throughput, latency, and memory utilisation, but performs inference purely on the CPU. While both make valuable contributions, neither tests dedicated edge accelerators such as the Hailo-10H NPU or current-generation Jetson hardware; both rely exclusively on Ollama; and both evaluate against generic benchmarks rather than edge-specific metrics.
 
Abstreiter et al. \cite{abstreiter2025sometimes} further investigate on-device LLM inference across CPU- and GPU-based SBC platforms, evaluating inference speed, energy, and micro-architectural behaviour. However, the study is limited to two devices (Raspberry Pi 5 and Jetson Orin Nano) and does not consider dedicated edge NPUs or integrate findings into an edge multi-dimensional framework. Tummalapalli et al. \cite{tummalapalli2026llm} benchmark a quantised 1.5B model across mobile devices, GPUs, and a Hailo-10H NPU, focusing on throughput, power, and thermal behaviour under sustained inference. Their findings highlight thermal throttling and memory bandwidth as key constraints, but the study is restricted to a single model and does not explore the broader configuration space or SBC-focused trade-offs. In OT and industrial contexts, existing LLM applications such as PLC code generation \cite{tran2024generating} and maritime data enrichment \cite{huang2024automating} have relied on cloud-hosted endpoints, underscoring the need for localised inference on cost-effective SBC hardware.

\subsection{Research Gap}

As positioned in Table~\ref{tab:related-work}, this work benchmarks modern GPU and NPU edge accelerators using their supported inference paths and introduces metrics that consider throughput alongside energy and physical size. Runtime and hardware effects are coupled in this design; controlled comparisons with additional frameworks such as TensorRT-LLM and ONNX Runtime are therefore left to future work.

\section{Experimental Setup} \label{sect:s3}

\subsection{IoT-Suitable Hardware Platforms}

Experiments were conducted across four edge platforms in five configurations, representing a spectrum of compute capabilities as detailed in Table~\ref{tab:edge-ai-specs}. These configurations span CPU-only, NPU-accelerated, and GPU-accelerated inference, enabling comparative evaluation of heterogeneous edge compute architectures.
 
\subsection{Model Selection}
 
We evaluate a set of compact LLMs representative of current edge-deployable models, spanning parameter sizes from 0.5B to 3B:
 
\begin{itemize}
    \item DeepSeek-R1-Distill-Qwen (1.5B)
    \item Qwen 2.5 family (0.5B and 1.5B variants)
    \item Qwen 2.5 Instruct and Coder variants (1.5B)
    \item Llama 3.2 (1B and 3B)
\end{itemize}
 
These models were selected to reflect a range of architectures and training objectives (general-purpose, instruction-tuned, and code-specialised) at parameter scales suitable for edge deployment. Critically, all selected models are well supported by their respective vendors for edge inference.

\begin{table*}[!htbp]
\centering
\small
\caption{Hardware characteristics relevant to efficiency benchmarking.}
\label{tab:edge-ai-specs}
\resizebox{\textwidth}{!}{%
\begin{tabular}{
    p{2.2cm}   
    p{2.4cm}   
    p{1.6cm}   
    p{2.2cm}   
    p{0.9cm}   
    p{0.9cm}   
    p{1.8cm}   
    p{2.6cm}   
    p{2.0cm}   
    p{2.0cm}   
}
\toprule
\textbf{Device} & \textbf{Dimensions} & \textbf{Price (\$)} &
\multicolumn{4}{c}{\textbf{CPU}} &
\multicolumn{3}{c}{\textbf{AI Accelerator}} \\
\cmidrule(lr){4-7} \cmidrule(lr){8-10}
& & & \textbf{Name} & \textbf{Cores} & \textbf{Speed} & \textbf{Memory}
& \textbf{Name} & \textbf{Speed / TOPS} & \textbf{Memory} \\
\midrule
\textbf{M5Stack LLM (Module LLM, M140)} &
54.0 × 54.0 × 13.0\,mm & \$99.90 &
AX630C SoC & 2 & Up to 1.2GHz & 4\,GB LPDDR4 (1\,GB usable) &
AX630C NPU & 3.2\,TOPS (INT8) & \(\sim\)3\,GB accelerator RAM \\
\midrule
\addlinespace
\textbf{Raspberry Pi 5} &
85 × 56 × 17\,mm & \$222.75 &
Broadcom BCM2712 (Cortex‑A76) & 4 & 2.4\,GHz & 8\,GB LPDDR4X‑4267 &
— & — & — \\
\midrule
\addlinespace
\textbf{Raspberry Pi 5 + AI HAT+ 2} &
85 × 56 × 20\,mm & \$451.83 &
Broadcom BCM2712 (Cortex‑A76) & 4 & 2.4\,GHz & 8\,GB LPDDR4X‑4267 &
Hailo‑10H & 40\,TOPS (INT4) & 8\,GB accelerator RAM \\
\midrule
\addlinespace
\textbf{NVIDIA Jetson Orin Nano Super Dev Kit} &
100 × 79 × 21\,mm & \$304.78 &
Arm Cortex‑A78AE & 6 & Up to 1.7\,GHz & 8\,GB LPDDR5, 102\,GB/s BW &
NVIDIA Ampere - 1024 CUDA cores, 32 tensor cores & 67\,TOPS (INT8) & Shared 8\,GB LPDDR5 \\

\bottomrule
\end{tabular}%
}
\end{table*}

\subsection{Inference Configuration}
 
Three runtime configurations were used across the testbed:
\begin{itemize}
    \item \textbf{Native Ollama}: Ollama's default HTTP API (\texttt{:11434/api/chat}), used on the Raspberry Pi 5 (CPU) and Jetson Orin Nano (CPU and GPU).
    \item \textbf{Hailo Ollama}: Hailo's custom Ollama server implementation (\texttt{:8000/api/chat}) for NPU-accelerated inference on the Raspberry Pi 5 + AI HAT+.
    \item \textbf{M5Stack StackFlow}: Runtime for the AX630C NPU.
\end{itemize}
 
All Ollama-based configurations used streaming inference, where tokens are returned incrementally as they are generated. Models were executed using 4-bit quantisation (Q4\_K\_M) as the primary precision level. Each inference call used a fixed prompt (``Explain why the sky is blue in two or more paragraphs.'') with a generation length capped at 100 tokens via the \texttt{num\_predict} parameter. Decoding parameters were held constant across all platforms to ensure comparability. The reported results therefore characterise warm-state inference and exclude model loading, runtime initialisation, and cold-start latency.
 
\subsection{Evaluation \& Multi-Dimensional Composite Metrics}
 
Performance was evaluated using three primary metrics:
 
\begin{itemize}
    \item \textbf{Throughput (tokens/s)}: the rate of token generation during inference.
    \item \textbf{Time-to-first-token (TTFT)}: latency between prompt submission and generation of the first token.
    \item \textbf{Energy consumption (MJ/Mtok)}: energy required to generate one million tokens, capturing hardware efficiency.
\end{itemize}
 
These metrics jointly capture responsiveness, sustained generation performance, and energy efficiency.

Edge and IoT deployments impose simultaneous constraints across power, physical footprint, and performance that conventional single-axis LLM benchmarks do not capture. A drone may require minimal power draw to preserve flight time, a satellite ground station may have a desired form-factor limit to accommodate other peripherals. To capture these trade-offs, we define two composite metrics that normalise token throughput against the physical and energy constraints relevant to IoT-suitable hardware:

\begin{itemize}
    \item \textbf{Throughput density (Tps/m\textsuperscript{3})}: Token throughput normalised by external device volume, indicating performance under physical space constraints. 
    \item \textbf{Energy per million tokens (MJ/Mtok)}: Measured system energy normalised to one million generated tokens, indicating efficiency
    under operational energy constraints.
\end{itemize}

These are comparative indicators rather than complete deployment metrics. Tps/m\textsuperscript{3} excludes items such as power supplies, enclosures and cooling clearance, while MJ/Mtok depends on the workload and test conditions; neither directly captures thermal throttling, memory bandwidth, or cold-start costs.
 
\subsection{Measurement Methodology}
 
Benchmarking was automated using a custom Python harness that interfaces with each runtime's HTTP API\footnote{Source code is publicly available \url{https://github.com/kaze-technologies/LLM-Edge-Benchmarking-Suite}}. For each model and hardware configuration, a warmup request is issued to ensure the model is loaded into memory before timed runs begin. The benchmark then issues a streaming inference request, recording wall-clock time at submission and at receipt of the first token (yielding TTFT), with throughput calculated as tokens generated divided by total elapsed time. Each configuration was executed for $n=5$ runs, with the mean reported and per-run values retained. Input power was measured using a Mecheer JK-PM07. Idle power was not subtracted, so the reported values represent total system
energy. Energy consumption was calculated as $\bar{P}t/N$ MJ/Mtok, where
$\bar{P}$ is mean measured power in watts, $t$ is the measured inference
duration in seconds, and $N$ is the number of generated tokens; numerically,
$1$ J/token equals $1$ MJ/Mtok.

\section{Results \& Analysis} \label{sect:s4}

The salient results\footnote{The full set of benchmarking results across all models and configurations is publicly available online \url{https://osf.io/5r9t4/overview}} are presented in Table~\ref{tab:benchmarking-results}. The experimental design enables comparison across hardware classes (CPU vs NPU vs GPU), model scales (0.5B–3B), and efficiency-constraint trade-offs.

\begin{table*}[!htbp]
\centering
\caption{LLM inference benchmarking results across edge hardware configurations}
\label{tab:benchmarking-results}
\renewcommand{\arraystretch}{1.3}
\setlength{\tabcolsep}{4pt}
\resizebox{\textwidth}{!}{%
\begin{tabular}{cl l ccccc}
\toprule
\textbf{Model} & \textbf{Size} & \textbf{Metric} & \textbf{M5Stack LLM} & \textbf{RPi 5} & \textbf{RPi5+HAT+} & \textbf{Jetson Orin CPU} & \textbf{Jetson Orin GPU} \\
\midrule

\multirow{3}{*}{DeepSeek-R1-Distill-Qwen-1.5B}
 & \multirow{3}{*}{1.5B}
   & Throughput (tok/s)              & 2.42 & 0.32 & 1.53 & 6.01 & 9.59\\
 & & Time-to-first-token (ms)        & 2.15 & 20.79 & 10.47 & 2.08 & 1.38\\
 & & Energy Consumption (MJ/Mtok) & 0.57 & 33.24 & 3.47 & 2.09 & 1.15\\
 \midrule

  \multirow{6}{*}{Qwen 2.5}
 & \multirow{3}{*}{0.5B}
    & Throughput (tok/s)              & -- & 0.86 & -- & 13.04 & 13.31\\
 & & Time-to-first-token (ms)        & -- & 8.96 & -- & 1.51 & 1.43\\
 & & Energy Consumption (MJ/Mtok) & -- & 10.83 & -- & 0.92 & 1.20\\
 \cmidrule{2-8}
 & \multirow{3}{*}{1.5B}
   & Throughput (tok/s)              & -- & 0.27 & 4.37 & 4.21 & 9.37\\
 & & Time-to-first-token (ms)        & -- & 29.98 & 2.36 & 4.74 & 2.13\\
 & & Energy Consumption (MJ/Mtok) & -- & 35.30 & 1.30 & 1.38 & 1.11\\
\midrule

   \multirow{3}{*}{qwen2.5:1.5b-instruct}
 & \multirow{3}{*}{1.5B}
   & Throughput (tok/s)              & 2.55 & 0.3 & 6.34 & 6.9 & 9.09\\
 & & Time-to-first-token (ms)        & 2.11 & 26.44 & 0.55 & 2.86 & 2.20\\
 & & Energy Consumption (MJ/Mtok) & 0.57 & 35.31 & 0.88 & 1.83 & 1.17\\
 \midrule

    \multirow{3}{*}{qwen2.5-coder:1.5b}
 & \multirow{3}{*}{1.5B}
   & Throughput (tok/s)              & -- & 0.3 & 2.07 & 6.76 & 8.66\\
 & & Time-to-first-token (ms)        & -- & 26.90 & 7.43 & 2.95 & 2.30\\
 & & Energy Consumption (MJ/Mtok) & -- & 36.01 & 2.57 & 1.87 & 1.16\\
 \midrule

\multirow{6}{*}{Llama 3.2}
  & \multirow{3}{*}{1B}
   & Throughput (tok/s)              & 3.44 & 0.39 & -- & 9.51 & 10.96\\
 & & Time-to-first-token (ms)        & 1.78 & 19.93 & -- & 2.09 & 1.82\\
 & & Energy Consumption (MJ/Mtok)        & 0.41 & 27.87 & -- & 1.64 & 1.23\\
 \cmidrule{2-8}
 & \multirow{3}{*}{3B}
   & Throughput (tok/s)              & -- & 0.14 & 1.01 & 4.3 & 6.31\\
 & & Time-to-first-token (ms)        & -- & 58.71 & 13.11 & 3.92 & 3.16\\
 & & Energy Consumption (MJ/Mtok) & -- & 76.64 & 5.51 & 3.04 & 1.72\\

\bottomrule
\end{tabular}
}

{\scriptsize Results are mean of $n=5$ runs. Generation length fixed at 100 tokens.\\-- = model not supported for device. MJ/Mtok = Energy consumption per million tokens (Mtok) in MegaJoules (MJ)}
\end{table*}

\subsection{Impact of Hardware Acceleration on Inference Efficiency}

\begin{figure*}[!htbp]
    \centering
    \includegraphics[width=\textwidth]{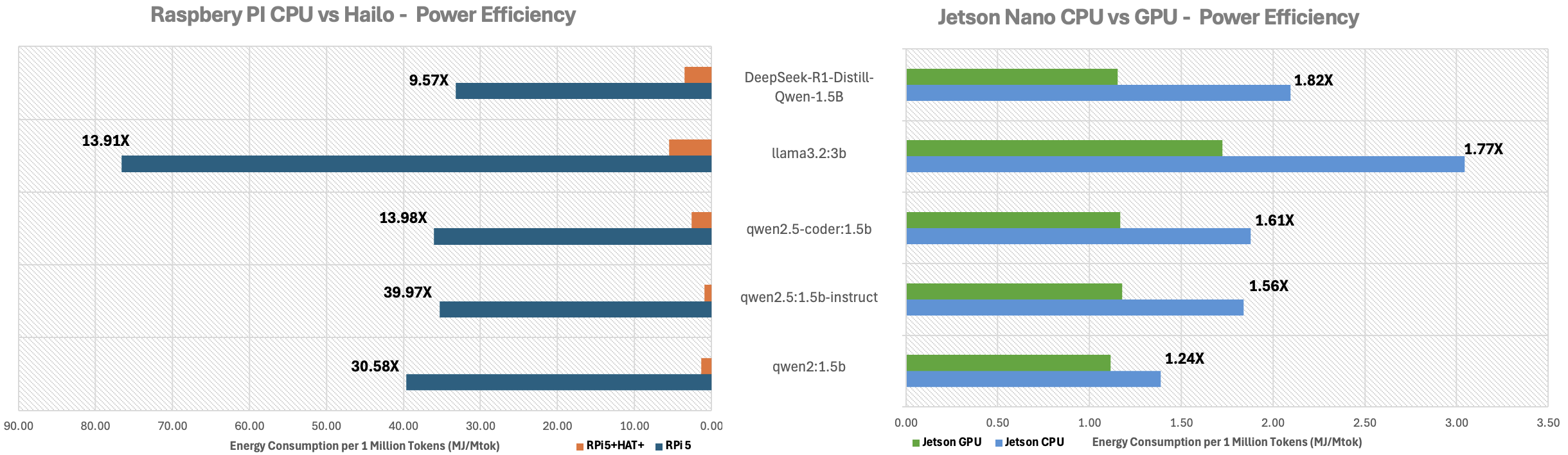}
    \caption{Comparison of power efficiency between CPU and available hardware accelerators on the Raspberry Pi 5 (left) and the Jetson Nano (right).}
    \label{fig:cpu-vs-accel}
\end{figure*}

Figure~\ref{fig:cpu-vs-accel} compares energy efficiency when offloading LLM inference from the CPU to a dedicated accelerator on both the Raspberry Pi 5 (Hailo-10H NPU) and Jetson Orin Nano Super (Ampere GPU). On the RPi 5, the Hailo NPU delivers 9.57$\times$ to 39.97$\times$ energy efficiency gains, reducing consumption from 27--77 MJ/Mtok (CPU) to 0.88--5.51 MJ/Mtok, while raising throughput from 0.14--0.86 tok/s to 1.01--6.34 tok/s.

On the Jetson, GPU offloading provides a more modest 1.24$\times$ to 1.82$\times$ efficiency gain, reflecting its stronger CPU baseline. In both cases, the accelerator also frees 
the host CPU for concurrent tasks such as sensor acquisition or network communication, a critical consideration in SBS deployments where the SBC serves multiple roles.

\subsection{Physical and Power Constraints for IoT Deployment}

\begin{figure}[!htbp]
    \centering
    \includegraphics[width=\columnwidth]{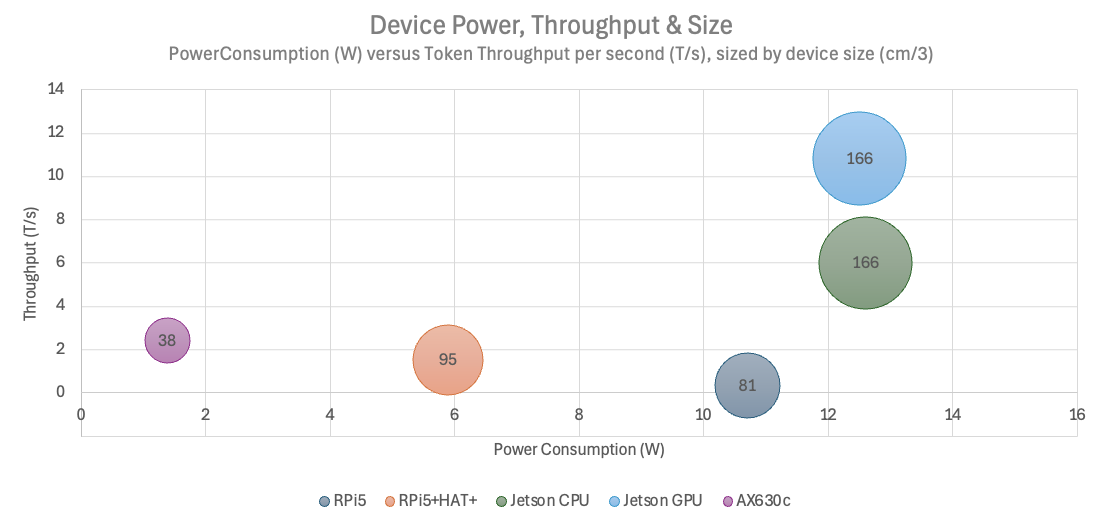}
    \caption{Device power consumption (W) against token throughput per second (T/s) where bubbles are sized according to the volume of the device (cm\textsuperscript{3}).}
    \label{fig:power-size}
\end{figure}

Figure~\ref{fig:power-size} plots power consumption against throughput for each configuration, with bubble size representing physical device volume (cm\textsuperscript{3}). The Jetson Orin GPU \& CPU configurations deliver the highest throughput but at 12--13W and 166 cm\textsuperscript{3}. The AX630c occupies the opposite corner: the smallest device (38cm\textsuperscript{3}) with lowest power draw (\textasciitilde 1.4W), achieving 2--3 tok/s. The RPi5+HAT+ is between these at ~6W and 95 cm\textsuperscript{3},  while the RPi 5 CPU-only configuration draws the most power relative to its output, consuming ~11W for under 0.4 tok/s. 

Notably, the RPi5+HAT+ delivers comparable throughput to the Jetson CPU at roughly half the power and in a smaller form factor, making it a strong candidate for power and space constrained deployments where the Jetson's GPU performance is not required.

\subsection{Multi-Dimensional Benchmarking Results}

\begin{figure*}[!htbp]
\centering
\begin{minipage}[t]{0.32\textwidth}
    \centering
    \includegraphics[width=\textwidth]{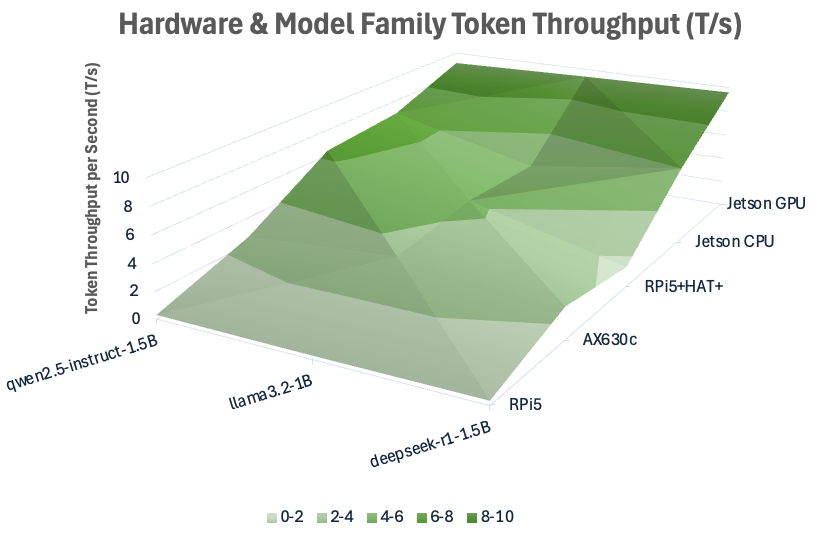}
    \subcaption{Token throughput (T/s).}
    \label{fig:throughput}
\end{minipage}
\hfill
\begin{minipage}[t]{0.32\textwidth}
    \centering
    \includegraphics[width=\textwidth]{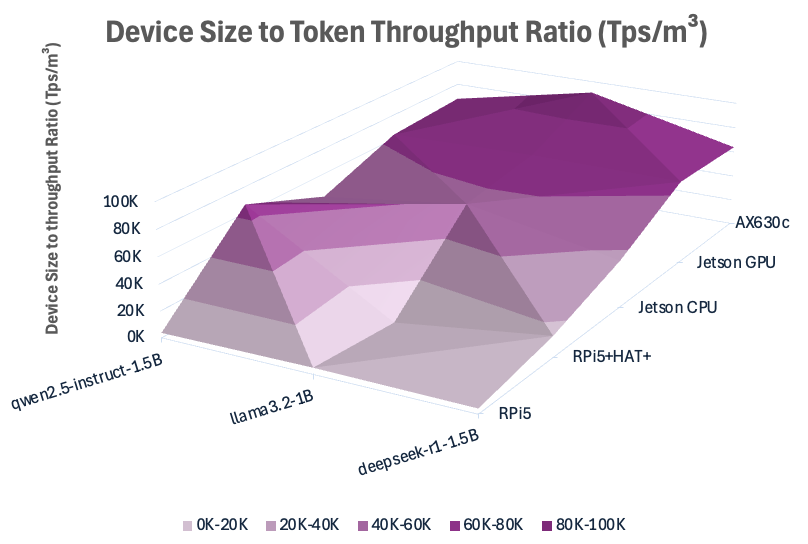}
    \subcaption{Volume-to-throughput ratio (Tps/m\textsuperscript{3}).}
    \label{fig:volume-ratio}
\end{minipage}
\hfill
\begin{minipage}[t]{0.32\textwidth}
    \centering
    \includegraphics[width=\textwidth]{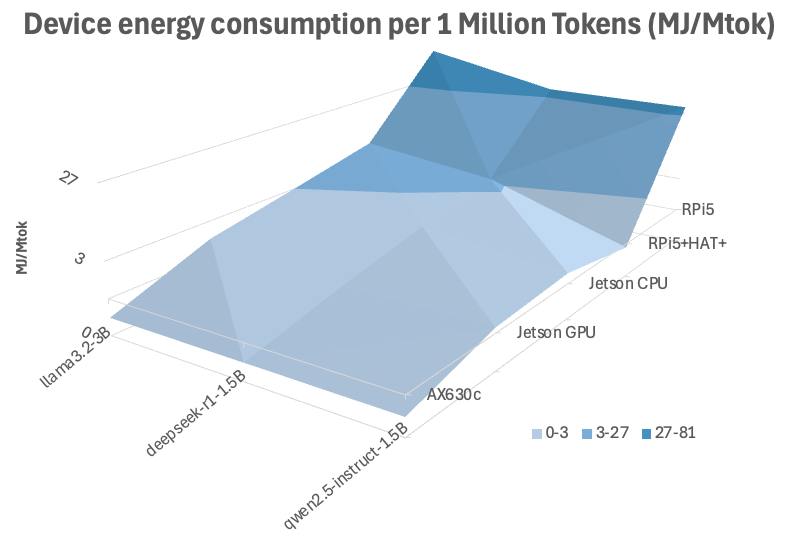}
    \subcaption{Energy consumption (MJ/Mtok).}
    \label{fig:energy-surface}
\end{minipage}
\caption{Multi-dimensional throughput ratios across hardware configurations for all shared supported LLMs: (a) raw token throughput, (b) throughput normalised by device volume, and (c) energy consumption per million tokens.}
\label{fig:throughput-ratios}
\end{figure*}

To evaluate edge LLM performance beyond raw speed, we introduce multi-dimensional throughput ratios that account for physical form factor and energy cost alongside tokens per second. Figure~\ref{fig:throughput-ratios} presents three surface plots across devices and the three models supported by all configurations (qwen2.5-instruct-1.5B, llama3.2-1B, deepseek-r1-1.5B).

Raw throughput (Figure~\ref{fig:throughput-ratios}a) follows an expected hierarchy: the Jetson Orin GPU leads at 9--10 tok/s, followed by Jetson CPU (4--7 tok/s), the RPi5+HAT+ and M5Stack AX630c (2--6 tok/s), and RPi 5 CPU trailing at under 0.4 tok/s. However, when throughput is normalised by device volume (Figure~\ref{fig:throughput-ratios}b), the M5Stack AX630c dominates at up to 90K+ Tps/m\textsuperscript{3}, owing to its 54$\times$54$\times$13mm form factor. This 
metric is relevant for space-constrained deployments such as ruggedised enclosures, wearable systems, or embedded installations where physical footprint is a primary design constraint.

Energy consumption per million tokens (Figure~\ref{fig:throughput-ratios}c) reveals a different ranking. The RPi 5 CPU is an order of magnitude worse than all accelerated platforms (33--76 MJ/Mtok vs 0.6--5.5 MJ/Mtok). Among accelerated configurations, the M5Stack and Jetson GPU are the most energy-efficient at 0.5--1.7 MJ/Mtok, while the RPi5+HAT+ varies more widely by model (0.8--5.5 MJ/Mtok), suggesting that NPU efficiency is possibly more sensitive to model architecture than GPU-based inference.

\section{Discussion: Implications for IoT in Critical Industries} \label{sect:s5}

\subsection{Resilient Distributed Cyber-Physical Systems}

The emergence of power-efficient edge AI hardware is fundamentally reshaping the design space for distributed AI in autonomous systems and contested environments. Compared to centralised infrastructure, which can require 300--380W or more per GPU, edge platforms operating in the 5--25W range enable scalable deployment across many distributed nodes, while also being mindful of privacy considerations and data contamination. This reduction in power consumption, combined with hardware acceleration, allows for efficient on-device inference while freeing the host CPU for concurrent tasks such as sensor processing, communications, and control. This may support local operation during connectivity loss, although system-level resilience was not evaluated here. This becomes increasingly important in environments with multi-level security data streams or nodes. 

\subsection{Energy-Constrained Autonomous Systems}

These characteristics are particularly impactful in autonomous and defence contexts, such as drone operations and distributed sensing environments. With constrained onboard energy budgets (e.g., 90--260Wh/kg \cite{PATTANAYAK2025126356}), drones benefit significantly from efficient inference that minimises power draw while enabling local decision-making and real-time data summarisation. This reduces the need for continuous high-bandwidth communication with centralised command systems, lowering both latency and detectability in contested environments. Local inference may reduce dependence on continuous connectivity, but resilience and trust are system-level properties not evaluated by these single-device experiments. Future comparisons of centralised, local-only and federated approaches should measure communication cost, energy, task performance, and robustness to compromised nodes.

\subsection{Distributed Satellite Ground Stations}

SBCs are becoming widely used for deploying portable, 
cost-effective satellite ground stations across distributed geographic locations \cite{10.1145/3730567.3764496}, with the Raspberry Pi particularly established as an accessible platform for open-source ground station implementations \cite{crian2024design}. Our results demonstrate that 1.5B parameter LLMs can run on these platforms at 2--10 tok/s with energy consumption as low as 0.57 MJ/Mtok, with a strong design incentive to use a hardware accelerator such as the Hailo-10H NPU for improved power efficiency, throughput, and to free the host CPU for concurrent ground station tasks such as antenna tracking, signal demodulation, and telemetry logging. Embedding LLM reasoning directly on the ground station enables a degree of intelligent autonomy: parsing and interpreting incoming commands, reasoning over sensor telemetry and system state to make best-available decisions, and adapting station behaviour during narrow satellite communication windows without depending on instructions from a centralised facility.

\section{Conclusion} \label{sect:s6}

This paper presented initial steps toward a multi-dimensional evaluation framework for LLM inference on IoT-suitable edge hardware, introducing two composite metrics: throughput density (Tps/m\textsuperscript{3}) and energy per million tokens (MJ/Mtok). These metrics extend conventional single-axis benchmarking to account for the physical and energy constraints of edge deployment. Benchmarking across four platforms revealed that hardware accelerators can deliver up to 40$\times$ energy efficiency gains over CPU-only inference. The M5Stack AX630c achieves the highest throughput density for space-constrained applications, while the Jetson Orin Nano GPU offers the best absolute throughput and the RPi 5 + Hailo-10H provides a strong balance of efficiency and form factor. These findings have direct implications for Internet of Everything deployments in critical industries, from resilient distributed cyber-physical systems to energy-constrained autonomous platforms and satellite ground stations. Future work will examine sparse and mixed-precision inference, additional runtimes including TensorRT-LLM and ONNX Runtime, cold-start and model-loading latency, sustained thermal and memory effects, and system-level comparisons of centralised and federated deployment.

\textbf{Availability of Data and Materials:} The data supporting the findings of this study are openly available from the Open Science Framework at \url{https://osf.io/5r9t4}.

\textbf{Ethics Approval:} Not applicable.

\textbf{Conflicts of Interest:} The authors declare no conflicts of interest to report regarding the present study.

\bibliographystyle{unsrt}  
\bibliography{references}

@article{liagkou5406285taming,
  title={Taming the LLMaaS Market: A Decision-Making Framework Utilizing Diverse Enterprise-Critical Selection Factors},
  author={Liagkou, Vasiliki and Fragiadakis, George and Filiopoulou, Evangelia and Nikolaidou, Mara and Michalakelis, Christos},
  journal={Available at SSRN 5406285},
  year={2025}
}

@article{zheng2025review,
  title={A review on edge large language models: Design, execution, and applications},
  author={Zheng, Yue and Chen, Yuhao and Qian, Bin and Shi, Xiufang and Shu, Yuanchao and Chen, Jiming},
  journal={ACM Computing Surveys},
  volume={57},
  number={8},
  pages={1--35},
  year={2025},
  publisher={ACM New York, NY}
}

@article{vaswani2017attention,
  title={Attention is all you need},
  author={Vaswani, Ashish and Shazeer, Noam and Parmar, Niki and Uszkoreit, Jakob and Jones, Llion and Gomez, Aidan N and Kaiser, {\L}ukasz and Polosukhin, Illia},
  journal={Advances in neural information processing systems},
  volume={30},
  year={2017}
}

@inproceedings{zhanginterpreting,
  title={Interpreting and improving large language models in arithmetic calculation},
  author={Zhang, Wei and Wan, Chaoqun and Zhang, Yonggang and Cheung, Yiu-ming and Tian, Xinmei and Shen, Xu and Ye, Jieping},
  booktitle={Proceedings of the 41st International Conference on Machine Learning},
  pages={59932--59950},
  year={2024}
}

@article{gomez2023confederacy,
  title={A Confederacy of Models: a Comprehensive Evaluation of LLMs on Creative Writing},
  author={G{\'o}mez-Rodr{\'\i}guez, Carlos and Williams, Paul and Glasbergen, Bree},
  journal={Findings of the Association for Computational Linguistics: EMNLP 2023},
  pages={14504--14528},
  year={2023},
  publisher={Association for Computational Linguistics}
}

@article{brin2023large,
  title={How large language models perform on the United States medical licensing examination: a systematic review},
  author={Brin, Dana and Sorin, Vera and Konen, Eli and Nadkarni, Girish and Glicksberg, Benjamin S and Klang, Eyal},
  journal={MedRxiv},
  pages={2023--09},
  year={2023},
  publisher={Cold Spring Harbor Laboratory Press}
}

@article{katz2024gpt,
  title={Gpt-4 passes the bar exam},
  author={Katz, Daniel Martin and Bommarito, Michael James and Gao, Shang and Arredondo, Pablo},
  journal={Philosophical Transactions of the Royal Society A},
  volume={382},
  number={2270},
  pages={20230254},
  year={2024},
  publisher={The Royal Society}
}

@inproceedings{huang2025llms,
  title={Llms at the edge: Performance and efficiency evaluation with ollama on diverse hardware},
  author={Huang, Donghao and Wang, Zhaoxia},
  booktitle={2025 International Joint Conference on Neural Networks (IJCNN)},
  pages={1--8},
  year={2025},
  organization={IEEE}
}

@article{grattafiori2024llama,
  title={The llama 3 herd of models},
  author={Grattafiori, Aaron and Dubey, Abhimanyu and Jauhri, Abhinav and Pandey, Abhinav and Kadian, Abhishek and Al-Dahle, Ahmad and Letman, Aiesha and Mathur, Akhil and Schelten, Alan and Vaughan, Alex and others},
  journal={arXiv preprint arXiv:2407.21783},
  year={2024}
}

@article{qwen2024qwen2,
  title={Qwen2. 5 technical report},
  author={Qwen, A Yang and Yang, Baosong and Zhang, Beichen and Hui, Binyuan and Zheng, Bo and Yu, Bowen and Li, Chengpeng and Liu, Dayiheng and Huang, Fei and Wei, Haoran and others},
  journal={arXiv preprint},
  year={2024}
}

@inproceedings{dettmers20218,
  title={8-bit Optimizers via Block-wise Quantization},
  author={Dettmers, Tim and Lewis, Mike and Shleifer, Sam and Zettlemoyer, Luke},
  booktitle={International Conference on Learning Representations}
}

@article{shi2025systematic,
  title={Systematic Characterization of LLM Quantization: A Performance, Energy, and Quality Perspective},
  author={Shi, Tianyao and Ding, Yi},
  journal={arXiv preprint arXiv:2508.16712},
  year={2025}
}

@misc{baller2021deepedgebenchbenchmarkingdeepneural,
      title={DeepEdgeBench: Benchmarking Deep Neural Networks on Edge Devices}, 
      author={Stephan Patrick Baller and Anshul Jindal and Mohak Chadha and Michael Gerndt},
      year={2021},
      eprint={2108.09457},
      archivePrefix={arXiv},
      primaryClass={cs.AI},
      url={https://arxiv.org/abs/2108.09457}, 
}

@article{abdulkadhim2026introducing,
  title={Introducing LEAF: LLM Edge Assessment Framework for Generative AI on the Edge},
  author={Abdulkadhim, Mustafa and Repas, Sandor R},
  journal={Machine Learning and Knowledge Extraction},
  volume={8},
  number={2},
  pages={48},
  year={2026},
  publisher={MDPI}
}

@article{nezami2025descriptor,
  title={Descriptor: Benchmark dataset for generative AI on edge devices (BeDGED)},
  author={Nezami, Zeinab and Hafeez, Maryam and Djemame, Karim and Zaidi, Syed Ali Raza and Xu, Jie},
  journal={IEEE Data Descriptions},
  year={2025},
  publisher={IEEE}
}

@article{gill2025edge,
  title={Edge AI: A taxonomy, systematic review and future directions},
  author={Gill, Sukhpal Singh and Golec, Muhammed and Hu, Jianmin and Xu, Minxian and Du, Junhui and Wu, Huaming and Walia, Guneet Kaur and Murugesan, Subramaniam Subramanian and Ali, Babar and Kumar, Mohit and others},
  journal={Cluster Computing},
  volume={28},
  number={1},
  pages={18},
  year={2025},
  publisher={Springer}
}

@misc{
lee2025qrazor,
title={{QR}azor: Reliable and Effortless 4-bit {LLM} Quantization by Significant Data Razoring},
author={Dongyoung Lee and Seungkyu Choi and Ik Joon Chang},
year={2025},
url={https://openreview.net/forum?id=lwcnZmyojm}
}

@misc{
picovoice2026Sub-4-Bit,
title={Sub-4-Bit LLM Quantization: Enterprise Guide to Model Compression \& Accuracy Tradeoffs},
author={picovoice},
year={2026},
url={https://picovoice.ai/blog/sub-4-bit-llm-quantization/}
}

@article{tummalapalli2026llm,
  title={LLM Inference at the Edge: Mobile, NPU, and GPU Performance Efficiency Trade-offs Under Sustained Load},
  author={Tummalapalli, Pranay and Arayakandy, Sahil and Pal, Ritam and Kundan, Kautuk},
  journal={arXiv preprint arXiv:2603.23640},
  year={2026}
}

@article{abstreiter2025sometimes,
  title={Sometimes painful but certainly promising: Feasibility and trade-offs of language model inference at the edge},
  author={Abstreiter, Maximilian and Tarkoma, Sasu and Morabito, Roberto},
  journal={arXiv preprint arXiv:2503.09114},
  year={2025}
}

@inproceedings{tran2024generating,
  title={Generating plc code with universal large language models},
  author={Tran, Kilian and Zhang, Jingxi and Pfeiffer, J{\'e}r{\^o}me and Wortmann, Andreas and Wiesmayr, Bianca},
  booktitle={2024 IEEE 29th International Conference on Emerging Technologies and Factory Automation (ETFA)},
  pages={1--8},
  year={2024},
  organization={IEEE}
}

@inproceedings{huang2024automating,
  title={Automating maritime risk data collection and identification leveraging large language models},
  author={Huang, Donghao and Fu, Xiuju and Yin, Xiaofeng and Pen, Haibo and Wang, Zhaoxia},
  booktitle={2024 IEEE International Conference on Data Mining Workshops (ICDMW)},
  pages={433--439},
  year={2024},
  organization={IEEE}
}

@inproceedings{shanmugarasa2025sok,
  title={Sok: The privacy paradox of large language models: Advancements, privacy risks, and mitigation},
  author={Shanmugarasa, Yashothara and Ding, Ming and Arachchige, Chamikara Mahawaga and Rakotoarivelo, Thierry},
  booktitle={Proceedings of the 20th ACM Asia Conference on Computer and Communications Security},
  pages={425--441},
  year={2025}
}

@inproceedings{10.1145/3730567.3764496,
author = {Chai, Wenchang and Liu, Jinhong and Zhang, Ziyue and Xia, Xianjin and Zheng, Yuanqing and Hou, Ningning and Yang, Qiang and Chen, Weiwei and Gu, Tao},
title = {Satellite IoT in Practice: A First Measurement Study on Network Availability, Performance, and Costs},
year = {2025},
isbn = {9798400718601},
publisher = {Association for Computing Machinery},
address = {New York, NY, USA},
doi = {10.1145/3730567.3764496},
booktitle = {Proceedings of the 2025 ACM Internet Measurement Conference},
pages = {891–899},
numpages = {9},
location = {USA},
series = {IMC '25}
}

@article{crian2024design,
  title={DESIGN AND IMPLEMENTATION OF A FULL-DUPLEX GROUND STATION FOR THE QO-100 SATELLITE SYSTEM BASED ON SDR AND RASPBERRY PI},
  author={Crișan, Nicolae},
  journal={Acta Technica Napocensis},
  volume={64},
  number={2},
  pages={9--14},
  year={2024},
  publisher={Universitatea Tehnica Cluj-Napoca}
}

@article{PATTANAYAK2025126356,
title = {Battery technology for sustainable aviation: a review of current trends and future prospects},
journal = {Applied Energy},
volume = {397},
pages = {126356},
year = {2025},
issn = {0306-2619},
doi = {https://doi.org/10.1016/j.apenergy.2025.126356},
url = {https://www.sciencedirect.com/science/article/pii/S0306261925010864},
author = {Tavish Pattanayak and Dimitri Mavris}
}

\end{document}